\documentclass[11pt]{JHEP3}
\usepackage{palatino}
\usepackage{amsmath}\usepackage{amssymb}
\usepackage{graphicx,epsfig}

\newcommand{\C}{{\mathbb C}}

\author{Joakim Munkhammar\\ Studentstaden 23:230, 752 33,
Uppsala, Sweden\\
E-Mail: \email{joakim.munkhammar@gmail.com}}

\title{Canonical Relational Quantum Mechanics from
Information Theory}


\abstract{In this paper we construct a theory of quantum mechanics
based on Shannon information theory. We define a few principles
regarding information-based frames of reference, including
explicitly the concept of information covariance, and show how an
ensemble of all possible physical states can be setup on the basis
of the accessible information in the local frame of reference. In
the next step the Bayesian principle of maximum entropy is
utilized in order to constrain the dynamics. We then show, with
the aid of Lisi's universal action reservoir approach, that the
dynamics is equivalent to that of quantum mechanics. Thereby we
show that quantum mechanics emerges when classical physics is
subject to incomplete information. We also show that the proposed
theory is relational and that it in fact is a path integral
version of Rovelli's relational quantum mechanics. Furthermore we
give a discussion on the relation between the proposed theory and
quantum mechanics, in particular the role of observation and
correspondence to classical physics is addressed. In addition to
this we derive a general form of entropy associated with the
information covariance of the local reference frame. Finally we
give a discussion and some open problems.}

\maketitle

\begin{document}
\newpage

\begin{center}
\textit{"Information is the resolution of uncertainty."}
\end{center}

\hspace{270pt} - Claude Shannon

\vspace{10pt}
\section{Introduction}
Quantum mechanics constitutes a conceptual challenge as it defies
many pivotal classical concepts of physics. Despite this the
theoretical and experimental success of quantum mechanics is
unparallel \cite{Rovelli}. The intuitive construction of classical
mechanics and the perhaps counter-intuitive quantum mechanical
formulation of reality has thus amounted to a problem of
interpretation of quantum mechanics. The list of interpretations
of quantum mechanics is long, but perhaps the most common
interpretations are: Copenhagen \cite{Bohr}, consistent histories
\cite{Griffiths}, many worlds \cite{Everett} and relational
\cite{Rovelli}. Many of these approaches share most central
features of quantum mechanics, the difference is mainly the
philosophical context in which they are situated. The canonical
problem in these approaches is perhaps the counter-classic, and in
many peoples views, counter-intuitive principle that physics is
fundamentally based on probability. The perhaps strongest opponent
of a probability-based theory of physics was Einstein who
constructed several unsuccessful thought experiments and theories
in order to disprove the commonly accepted view of quantum
mechanics \cite{Einstein}. As a matter of fact all approaches to
create a non-probabilistic version of quantum mechanics have
failed \cite{Bransden}. It has also been proven, via for example
Bell's theorems, that the construction of such a deterministic or
"local hidden variable theory" is impossible \cite{Bransden}. Thus
one has no choice but to conclude that physics, inevitably, has
quantum properties.


\subsubsection{Relational Quantum Mechanics}
It was Einstein's revised concepts of simultaneity and frame of
reference that inspired Rovelli to formulate a theory of mechanics
as a relational theory; \textit{relational quantum mechanics}. In
this seminal approach frames of reference were utilized and only
relations between systems had meaning. This setup gave interesting
solutions to several quantum-related conceptual problems such as
the EPR-paradox \cite{Rovelli}. It may be concluded that Rovelli
recast physics in the local frame of reference in such a way that
the interactions between systems amounted to the observed quantum
phenomena. A particular facet of Rovelli's approach was that any
physical system could be observed in different states by different
observers "simultaneously". This property can be interpreted as an
extension of the concept of simultaneity in special relativity
\cite{Rovelli}. The theory of relational quantum mechanics was
based on the hypothesis that quantum mechanics ultimately arose
when there was a lack of information of investigated systems.

\subsubsection{Universal action reservoir}
In a recent paper Lisi gave an interesting approach to quantum
mechanics based on information theory and entropy \cite{Lisi}. He
showed that given a \textit{universal action reservoir} and a
principle for maximizing entropy quantum mechanics could be
obtained. In his paper the origin of the universal action
reservoir was postulated as a principle and was given no deeper
explanation. This was addressed in recent papers by Lee
\cite{Lee,Lee2} where he suggested that it was related to
information theory coupled to causal horizons.


\subsubsection{Relational quantum mechanics and the universal
action reservoir}
 In this paper we shall recast Rovelli's
relational approach to quantum mechanics on the concept of
information covariance and connect it to Lisi's canonical
information theoretic approach. In the process of this we show
that the universal action reservoir is an inevitable consequence
of incomplete information in physics. This theory is, in spirit of
Rovelli's approach, a generalization of the special relativistic
concept regarding frames of reference.

%

\section{Physics with information covariance}

\subsection{Principles for information-based frames of reference}
Let us assume that we have a set of observations of certain
physical quantities of a physical system. This is the obtained
information regarding the system. For all other events that has
not been observed we only know that \textit{something} possible
happened. If we knew with certainty \textit{what} happened then we
as observers would be in a frame of reference based on complete
information regarding the explored system. However such a theory
would require some form of a conservation law which would require
that a physical system has a predetermined state when not
observed. Let us instead consider the opposite: Assume that no
observation is made of the system, then information regarding that
system is not accessible and thus not inferable without more
observations. It is then reasonable that \textit{anything
physically possible} could have occurred when it was not observed.
If one can base the laws of physics on the premise of only what is
known in the frame of reference, in terms of information,
regarding any physical system in connection with the observer and
her frame then one has attained a high from of generality in the
formulation of the laws of physics. This amounts to a seemingly
tautological yet powerful principle:

\vspace{15pt}

\begin{center}
\textit{If a system in physics is not
observed it is in any physically possible state.}
\end{center}

\vspace{15pt}

We shall call this principle \textit{the principle of physical
ignorance}. In one sense this principle is intuitive and trivial:
the unknown is not known. In another more intricate sense it
violates a number of basic principles of physics such as many of
the laws of classical physics. The law of inertia is one such
principle for example; the inertia of a body does not necessarily
hold when we do not observe it. However one cannot assume that
Newton's laws or any other classical laws hold in a system for
which limited information is known. For all we know regarding
classical physics is that it holds in a certain classical limit.
The definition of "known" here is what information has become
accessible in ones frame of reference obtained through interaction
with another system. The laws of physics should thus be formulated
in such a way that they hold regardless of information content in
the local frame of reference; the formulation of the laws of
physics should be invariant with respect to the information
content. This amounts to a principle for the formulation of the
laws of physics:

\vspace{15pt}

\begin{center}
\textit{The laws of physics are defined on the basis of the
information in the frame of reference.}
\end{center}

\vspace{15pt}


This principle shall be call the \textit{principle of information
covariance}. It's scope of generality is similar to the
\textit{principle of general covariance}, which is the natural
generalization of the frames of reference used in special
relativity. The \textit{principle of information covariance}
suggests that the frame of reference is entirely based on
information; thus any new observations will alter the information
content and thus re-constrain the dynamics of the studied system
"projected" within the frame of reference. One should also keep in
mind that physics cannot, in a frame of incomplete information,
"project" any information through physical interactions that is
not inferable from the accessible information. This creates a form
of locality of physics: Physics only exists in every frame of
reference. Thus there exists no such thing as an objective
observer. It should be noted here that there exists a similar
notion of frames of reference in traditional relational quantum
mechanics \cite{Rovelli}. The dynamics is derived from a different
set of principles and formal setup but arrives at a similar
conclusions \cite{Rovelli}.


\subsection{Formal setup}
Let us assume that we have a set of observations $A_n$ regarding
physical quantities of a system $C$ from a frame of reference $K$.
The available information regarding the system $C$ in $K$ is given
by the information in the observed physical quantities $A_n$ and
what can be inferred from them, the rest of the properties of the
system are by \textit{the principle of physical ignorance}
unknown. Formally according to \textit{the principle of physical
ignorance} we may conclude that the possible configurations of the
explored physical system $C$ in the frame of reference $K$ has to
form a set of all possible physical configurations under the
constraint of the set of observations $A_n$ for the system.
Consider a set of configuration parameters for each possible
physical configuration or path in configuration space $path =
q(t)$. The set of configurations parameters is parameterized by
one or more parameters $t$. Then for each physical system we may
associate an action $S[path] = S[q(t)]$ defined on the basis of
the configuration space path $q(t)$. The structure of the action
is here assumed to be something of a universal quantity of
information inherent to every frame of reference, it is the paths
in the configuration space of objects that is unknown to every
observer until observed. The action of a system is defined
classically as $S = \int L(q,\dot{q})dt$, where $L(q,\dot{q})$ is
the Lagrangian for the system \cite{Thornton}. Traditionally with
the aid of a variation principle the expected action of a
classical system is used to derive the dynamics of the system
\cite{Thornton}. In quantum mechanics a probabilistic setup is
performed, ideally via a path integral formulation. In our
situation we shall instead look for all possible configurations
under condition of the observations $A_n$ in accordance with the
previously defined \textit{principle of physical ignorance}. The
possible actions, based on the possible configurations, are each
associated with a probability of occurring, $p[path] = p[q(t)]$.
The sum of these probabilities need to satisfy the ubiquitous
normalization criterion:

\begin{equation}\label{Normalization}
1 = \sum_{paths} p[path] = \int Dqp[q],
\end{equation}

Along with this we may conclude that any functional, or observable
quantity $Q$, has an expected value according to the expression
\cite{Lisi}:

\begin{equation}\label{Expected1}
\langle Q \rangle = \sum_{paths} p[path] Q[path] = \int Dq
p[q]Q[q].
\end{equation}

We have the expected action $\langle S \rangle$ according to
\cite{Lisi}:

\begin{equation}\label{ExpectedAction}
\langle S \rangle = \sum_{paths} p[path] S[path] = \int Dq
p[q]S[q].
\end{equation}

As a measure of the information content, or rather lack thereof,
we can construct the entropy of the system according to:

\begin{equation}\label{Entropy}
H = -\sum_{paths} p[path] \log p[path] = -\int Dq p[q]\log p[q].
\end{equation}

So far we have merely utilized information theoretic concepts, we
shall deal with its physical consequences further on in this
paper. Although the probabilities for the events in system $C$
observed from reference system $K$ are yet undefined we have a set
of possible configurations that could occur for a system and we
have associated a probability of occurrence with each based on the
ensemble setup above. In order to deduce the probability
associated with each possible event we need some form of
restriction on the ensemble of possibilities. In a Bayesian theory
of interference there is a maximization principle regarding the
entropy of a system called the \textit{Principle of maximum
entropy} \cite{Jaynes1} which postulates the following:

\vspace{10pt}

\begin{center}
\textit{Subject to known constraints, the probability distribution
which best represents the current state of knowledge is the one
with largest entropy.}
\end{center}

\vspace{10pt}

This principle is utilized in several fields of study, in
particular thermodynamics where it serves as the second
fundamental law \cite{Jaynes1}. We shall assume that this
principle holds and we shall utilize it as a restriction on our
framework. In \cite{Lisi} Lisi performed the following derivation
which is worth repeating here. By employing Lagrange multipliers,
$\lambda \in \C$ and $\alpha \in \C$, the entropy \eqref{Entropy}
is maximized by:

\begin{equation}
H'= - \int Dq p[q] \log p[q] + \lambda \Big(1-\int Dqp[q]\Big) +
\alpha \Big(\langle S \rangle - \int Dq p[q]S[q]\Big),
\end{equation}

which simplified becomes:

\begin{equation}
H' = \lambda + \alpha \langle S\rangle - \int Dq (p[q]\log p[q] +
\lambda p[q] + \alpha p[q] S[q]).
\end{equation}

If we perform variation on the probability distribution we get:

\begin{equation}
\delta H' = - \int Dq (\delta p[q])(\log p[q] + 1 + \lambda +
\alpha S[q])
\end{equation}

which is extremized when $\delta H' = 0$ which corresponds to the
probability distribution:

\begin{equation}\label{Probability}
p[q] = e^{-1-\lambda} e^{-\alpha S[q]} = \frac1Z e^{-\alpha S[q]}
\end{equation}

which is compatible with the knowledge constraints \cite{Lisi}. By
varying the Lagrange multipliers we enforce the two constraints,
giving $\lambda$ and $\alpha$. Especially one gets:
$e^{-1-\lambda} = \frac1Z$ where $Z$ is the partition function on
the form:

\begin{equation}\label{Partition}
Z = \sum_{paths} e^{-\alpha S[path]} = \int Dq e^{-\alpha S[q]}
\end{equation}

and the parameter $\alpha$ is determined by solving:

\begin{equation}
\langle S \rangle = \int Dq S[q]p[q] = \frac1Z \int Dq S[q]
e^{-\alpha S[q]} = - \frac{\partial}{\partial \alpha} \log Z.
\end{equation}

In order to fit the purpose Lisi concluded that the Lagrange
multiplier value; $\alpha \equiv \frac{1}{i\hbar}$. Lisi concluded
that this multiplier value was an intrinsic quantum variable
directly related to the average path action $\langle S \rangle$ of
what he called the universal action reservoir. In similarity with
Lisi's approach we shall also assume that the arbitrary
scaling-part of the constant $\alpha$ is in fact $1/\hbar$. Lisi
also noted that Planck's constant in $\alpha$ is analogous to the
thermodynamic temperature of a canonical ensemble, $i \hbar
\leftrightarrow k_B T$; being constant reflects its universal
nature - analogous to an isothermal canonical ensemble
\cite{Lisi}. This assumption along with \eqref{Partition} brings
us to the following partition function:

\begin{equation}\label{Partition15}
Z = \sum_{paths} e^{i\frac{S[path]}{\hbar}} = \int Dq
e^{i\frac{S[q]}{\hbar}}.
\end{equation}

By inserting \eqref{Partition15} into \eqref{Expected1} we arrive
at the following expectation value for any physical quantity $Q$:

\begin{equation}\label{ExpectedQuantity}
\langle Q\rangle = \sum_{paths} p[path] Q[path] = \int Dq Q[q]p[q]
= \frac1Z \int Dq Q[q] e^{i\frac{S[q]}{\hbar}},
\end{equation}


This suggests that a consequence of the incomplete information
regarding the studied system is that physics is inevitably based
on a probabilistic framework. Conversely, had physics not been
probabilistic in the situation of incomplete information then
information of the system could be inferred. But a process of
inferring results from existing limited information does not
provide more information regarding that system than the limited
information had already provided. That would have required, as we
previously argued, an additional principle of perfect information.
Instead it is only interaction that can provide new information.
We may conclude that by the \textit{principle of information
covariance} physics is local and based only on the available
information in the local information-based frame of reference. In
turn this this creates an ensemble of possible states with a
definite and assigned expectation value for each physical quantity
in the studied system according to \eqref{ExpectedQuantity}. This
formalism, which might be called \textit{information covariant},
is then directly compatible with the general principle of
relativity wherein \textit{All systems of reference are equivalent
with respect to the formulation of the fundamental laws of
physics}.




\begin{figure}
\begin{center}\label{fig1}
\epsfig{file=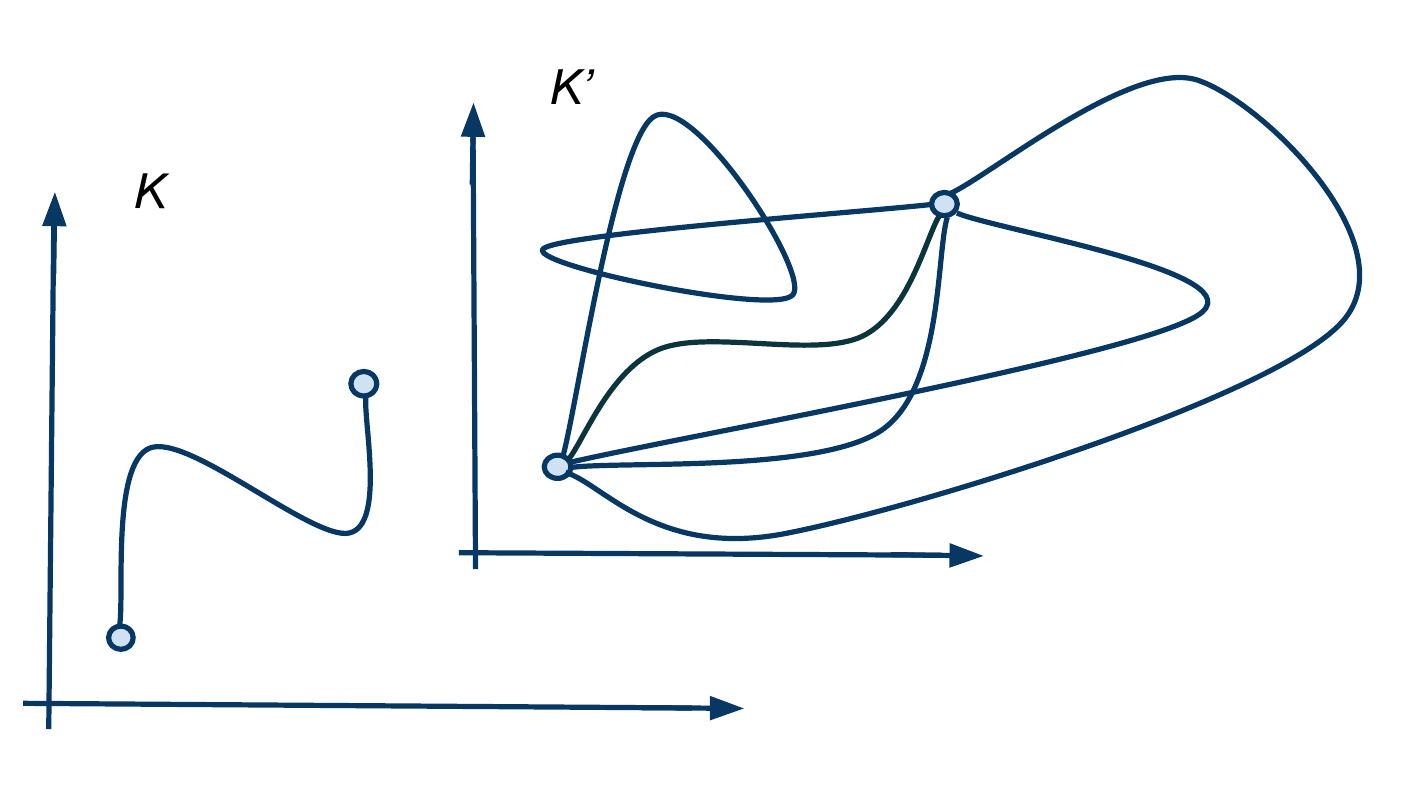,height=170pt,width=310pt}
\end{center}
 \caption{ This illustration shows the path of a particle from one
 point to another in a complete information frame of reference $K$
 (essentially a particle that is observed along its path). It also
shows some of the possible paths a particle takes in the
incomplete information frame of reference $K'$.}
\end{figure}

\section{Connections to quantum mechanics}
\subsection{Path integral formulation}\label{PathIntegral}
The path integral formulation, originally proposed by Dirac but
rigorously developed by Feynman \cite{Feynman}, is perhaps the
best foundational approach to quantum mechanics available
\cite{Lisi}. It shows that quantum mechanics can be obtained from
the following three postulates assuming a quantum evolution
between two fixed endpoints \cite{Feynman}:

\begin{itemize}
\item[\textbf{1.}] The probability for an event is given by the
squared length of a complex number called the probability
amplitude.

\item[\textbf{2.}] The probability amplitude is given by adding
together the contributions of all the histories in configuration
space.

\item[\textbf{3.}] The contribution of a history to the amplitude
is proportional to $e^{iS/\hbar}$, and can be set equal to $1$ by
choice of units, while $S$ is the action of that history, given by
the time integral of the Lagrangian $L$ along the corresponding
path.
\end{itemize}

In order to find the overall probability amplitude for a given
process then one adds up (or integrates) the amplitudes over
postulate $3$ \cite{Feynman}. In an attempt to link the concept of
information-based frames of reference - developed in this paper -
to quantum mechanics we shall utilize Lisi's approach wherein the
probability for the system to be on a specific path is evaluated
according to the following setup (see \cite{Lisi} for more
information). The probability for the system to be on a specific
path in a set of possible paths is:

\begin{equation}
p(set) = \sum_{paths} \delta_{path}^{set} p[path] = \int Dq
\delta(set-q)p[q].
\end{equation}

Here Lisi assumed that the action typically reverses sign under
inversion of the parameters of integration limits:

\begin{equation}
S^{t'} = \int^{t'} dt L(q,\dot{q}) = - \int_{t'} dt L(q,\dot{q}) =
- S_{t'}.
\end{equation}

This implies that the probability for the system to pass through
configuration $q'$ at parameter value $t'$ is:

\begin{align}\label{PsiProbability}
p(q',t') = \int Dq \delta(q(t') -q)p[q] =
\Bigg(\int^{q(t')=q'}
Dqp^{t'}[q] \Bigg) \Bigg(\int_{q(t')=q'}Dqp_{t'}[q] \Bigg) =
\psi(q',t')\psi^\dagger(q',t'),
\end{align}

in which we can identify the quantum wave function:

\begin{equation}
\psi = \int^{q(t')=q'} Dq p^{t'} [q] = \frac{1}{\sqrt{Z}}
\int^{q(t')=q'} Dq e^{-\alpha S^{t'}} = \frac{1}{\sqrt{Z}}
\int^{q(t')=q'} Dq e^{i\frac{S^{t'}}{\hbar}}.
\end{equation}

The quantum wave function $\psi(q',t)$ defined here is valid for
paths $t<t'$ meeting at $q'$ while its complex conjugate
$\psi^\dagger(q',t')$ is the amplitude of paths with $t>t'$
leaving from $q'$. Multiplied together they bring the probability
amplitudes that gives the probability of the system passing
through $q'(t')$, as is seen in \eqref{PsiProbability}. However,
just as Lisi points out \cite{Lisi}, this quantum wave function in
quantum mechanics is subordinate to the partition function
formulation since it only works when $t'$ is a physical parameter
and the system is $t'$ symmetric, providing a real partition
function $Z$. Indeed, the postulate of an information covariant
setup on the laws of physics according to the previous section
suggests that physics is ruled by the general complex partition
function \eqref{Partition}:

\begin{equation}\label{Partition2}
Z = \sum_{paths} e^{i\frac{S[path]}{\hbar}} = \int Dq e^{i
\frac{S[q]}{\hbar}}.
\end{equation}

How does this relate to the path integral formulation? The sum in
the partition function \eqref{Partition2} is a sum over paths. Let
us take the common situation when the path is that of a particle
between two points. We can then conclude that each term is on the
form $e^{iS[path]/\hbar}$ which is equivalent to postulate $3$.
Furthermore all paths are added, thus postulate $2$ is also
checked. Also, at least for the situation where $p(q',t')
=\psi(q',t')\psi^\dagger(q',t')$ the sum adds up to the
probability density, checking postulate $1$ as well. Thus we may
conclude that the information covariant approach is equivalent to
the canonical \textit{path integral formulation} of quantum
mechanics under the circumstances provided for it.

\begin{figure}
\begin{center}\label{fig2}
\epsfig{file=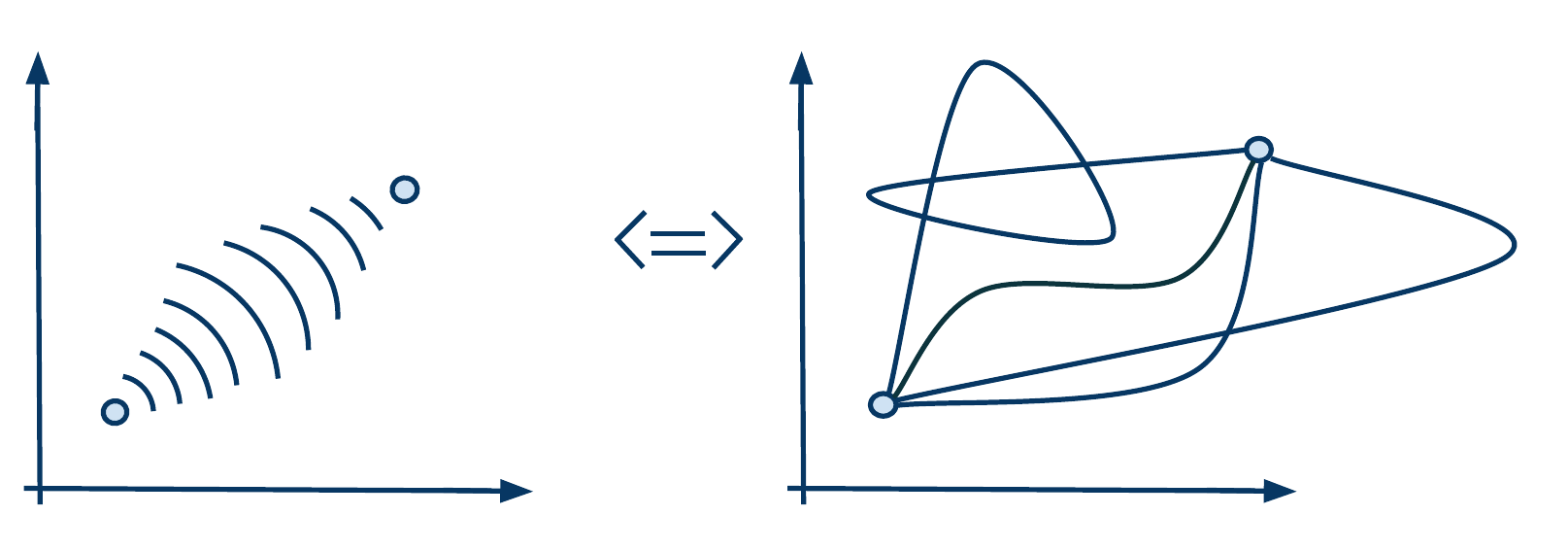,height=110pt,width=350pt}
\end{center}
\caption{This illustration shows on the left hand side the
uncertainty of path of a particle from one point to another and on
the right hand side that the particle takes all possible paths
from one point to another. These two interpretations are
equivalent under the general interpretation that information is
incomplete. Uncertainty in path means in practice that it takes
any possible path until we observe it, a superposition of states
is inevitable when information is incomplete.}
\end{figure}

\subsection{Quantum properties}
The path integral formulation is canonical for quantum mechanics
and covers the wide variety of special features inherent to
quantum mechanics \cite{Feynman,Lisi}. Since the approach in this
paper is equivalent to the path integral formulation in most
aspects, some properties are be worth discussing. A pivotal
component of quantum mechanics is the canonical commutation
relation which gives rise to the Heisenberg uncertainty principle
\cite{Bransden,Feynman}. For example the famous commutation
relation between position $x$ and momentum $p$ of a particle is
defined as:

\begin{equation}
[x,p] = i \hbar.
\end{equation}

This can be obtained through the path integral formulation by
assuming a random walk of the particle from starting point to end
point \cite{Feynman}. This works with this theory as well under
the same considerations since a random walk is equivalent to a
walk with no information about direction. In the path integral
formulation it is also possible to show that for a particle with
classical non-relativistic action (where where $m$ is mass and $x$
is position):

\begin{equation}
S = \int \frac{m \dot{x}^2}{2} dt,
\end{equation}

that the partition function $Z$ in the path integral formulation
turns out to satisfy the following equation \cite{Feynman}:

\begin{equation}\label{Schrodinger}
i \hbar \frac{\partial Z}{\partial t} = \Big[-\frac12 \nabla^2 +
V(x)\Big] Z.
\end{equation}

This is the Schr\"odinger equation for $Z=\psi$ and where $V(x)$
is a potential \cite{Bransden}. It is also possible to show the
conservation of probability from the Schr\"odinger equation
\eqref{Schrodinger} \cite{Bransden}. Here we can see that the
traditional usage of operators on a Hilbert space in quantum
mechanics is a useful tool when information is incomplete. Another
interesting aspect of quantum mechanics is the
\textit{superposition principle} which states that a particle
occupies all possible quantum states simultaneously
\cite{Bransden}. That the dynamics of a system is fundamentally
unknown or occupying all states simultaneously are both parts of
the same concept that information is incomplete regarding the
system. The popular quantum superposition thought experiment
\textit{Schr\"odingers cat} in which the alive/dead state of a cat
in a hazardous closed box is also evidently based on the lack of
information regarding the state of the cat. The superposition is
intuitively equivalent to the lack of information. The resolution
of this problem in this theory is that the state of the cat is
fundamentally unknown in our system of reference until we open the
box and observe it, thus obtaining information. The situation is
also relational because even when information is obtained the
information is only inherent to our frame of reference which might
not necessarily be the same for any other frame of reference. For
a complete verification that quantum mechanics can be interpreted
as a relational theory see \cite{Rovelli}. The proof that quantum
mechanics can be interpreted as relational is constructed with
Bra-kets in Hilbert spaces and is directly related to a linearity
inherent to quantum mechanics. Another interesting topic worth
mentioning here is the famous double slit experiment. The setup is
as follows: one has two slits and behind them a detection screen
is setup. Some form of beam of particles is then sent through the
slits and a statistical pattern is shown on the detection screen
\cite{Bransden}. The result is an interference pattern equivalent
to that described by the path integral formulation of quantum
mechanics. Such a pattern is not expected in classical mechanics.
The interpretation from the theory developed in this paper goes as
follows: due to the lack of information in the local
information-based frame of reference a particle takes any possible
path, a similar interpretation was given by Lee \cite{Lee}. This
situation is equivalent to the path integral formulation. The
particle is, in our frame of reference, wave-like until we observe
it. This shows how the ubiquitous wave-particle dualism arises
under the lack of information. As far as observation goes, we will
discuss that in section \ref{Observation} below. Quantum
entanglement is also a particular feature of quantum mechanics
that has spurred interpretational complications regarding quantum
mechanics. Quantum entanglement implies, among other things, that
information can travel at a superluminal speeds in most
interpretations of quantum mechanics \cite{Einstein}. This
violates the principle of invariant speed of light inherent to
special relativity \cite{Einstein}. However it has been solved for
relational quantum mechanics by postulating that physics is local,
and then it can be shown that no superluminal transfer of
information occurs \cite{Lee2}. Since the theory presented in this
paper by design is relational the same conceptual solution holds.
We will discuss the connection between the theory developed in
this paper and Rovelli's version of relational quantum mechanics
in section \ref{SectionRelational} below. Since the theory
developed in this paper does not violate quantum mechanics it
ought also to be completely compatible with the Bohm-De Broglie
pilot-wave approach \cite{Bohm} to quantum mechanics.

%


\subsection{Connections to relational quantum
mechanics}\label{SectionRelational} Relational quantum mechanics
is a theory of quantum mechanics based on the notion that only
systems in relation have meaning \cite{Rovelli}. The observer and
the partially observed system makes out such a system typically.
This addresses the \textit{problem of the third person} or
\textit{Wigner's friend} as it is also called in which an observer
observes another observer observing. The problem is solved by
assuming that the two observers may "simultaneously" observe
different states regarding the same object under observation. This
is shown to be a legal construction in quantum mechanics and is,
as Rovelli points out, non-antagonistic towards the most common
formulations of quantum mechanics \cite{Rovelli}. Furthermore, in
his seminal paper introducing relational quantum mechanics Rovelli
postulates that quantum mechanics arises from the lack of
information in classical physics. Given these facts one might ask
the following question: what is the similarity and what is the
difference between Rovelli's approach and the theory developed in
this paper? First of all our approach is relational and based on a
local frame of reference, which is similar to the concept used by
Rovelli. Second, this theory of quantum mechanics is based on
information which is similar to that used in Rovelli's approach.
There are two main differences. First of all this theory is based
on a few principles - different than those used by Rovelli - that
sets the foundation for a information covariant relational theory
of physics. Second, it utilizes an ensemble setup of Shannon
information theory, developed by Lisi, which equips relational
quantum mechanics with an information-based path integral
formulation.


\subsection{Correspondence principle}
The correspondence from quantum mechanics, or any quantum field
theory, to classical physics is when $\hbar \to 0$ or more
generally when $S>>\hbar$. Since this theory is equivalent to the
canonical path integral approach to quantum mechanics under
reasonable considerations we may state tautologically that the
correspondence to classical physics follows the same limits as for
regular quantum mechanics. The meaning of the correspondence is
also reasonable: If $\hbar$ descends to zero the partition
function will "collapse" and give only one expected value for each
quantity: The most expected one which by the Ehrenfest theorem is
the classical \cite{Bransden}. If the action is very large
($S>>\hbar$) the situation is the same; the larger the ratio
between the classical action and Planck's reduced constant is the
more likely the classical outcome is.

\subsection{Observations and wave function
collapse}\label{Observation}

Observation is by definition obtaining information from a studied
object \cite{Munkhammar,Lisi}. In order to obtain information
regarding a system one has to interact with it. This suggests that
observation of a system in practice is interacting with it, a view
of observation that is also held within the field of relational
quantum mechanics \cite{Rovelli}. In quantum mechanical terms when
an observation is performed then a wave function collapse occurs
\cite{Bransden}. In the theory developed in this paper the
probability for a specific path (or state) becomes one. Naturally
the quantum expectation value for a quantity $Q$ simply becomes
the one for the observed path $A$:

\begin{equation}
\langle Q \rangle = \sum_{paths} p[path] Q[path] = \int Dq
p[q]Q[q] = \frac1Z \int Dq Q[q] e^{i\frac{S[q]}{\hbar}} = Q[A].
\end{equation}

Observation of a system limits the possibilities of that system by
obtaining information about it. The kinematics of a system, as
viewed from our frame of reference, is based on the local
information about it by \textit{the principle of information
covariance}.



\section{Notes on relativistic invariance}
The theory developed in this paper is by definition implicitly
relativistic. The relativistic kinematics inherent to special
relativity should hold in this theory under at least the same
conditions for which the canonical path integral formulation is
relativistic. In the situation of complete information (or
$S>>\hbar$) the laws of (special) relativity hold in the classical
sense. We shall not detail the relativistic concepts further in
this paper, nor shall we attempt at constructing a general
relativistic, or a \textit{general information covariant} theory
of gravitation for the information-based frames of reference.
Instead, this is left for future investigation. However one might
presume that such an approach might share certain properties with
the relational approach to quantum gravity \cite{Rovelli}.

\section{Entropy}\label{SectionEntropy}
A great deal of this paper has consisted of meshing together
mathematical structures derived by previous authors under a new
set of principles. In contrast to this we shall here provide an
explicit calculation of the entropy associated with an
information-based frame of reference. We defined the entropy as
follows \eqref{Entropy}:

\begin{equation}\label{Entropy2}
H = -\sum_{paths} p[path] \log p[path] = -\int Dq p[q]\log p[q].
\end{equation}

The entropy \eqref{Entropy2} is based purely on information theory
and has thus no obvious direct connection to physical quantities.
Let us for this sake allow a scaling constant between the
information entropy $H$ and the thermodynamical entropy
$\mathcal{H}$:

\begin{equation}
\mathcal{H} \equiv k H.
\end{equation}

It was shown that after maximizing the entropy the probability of
a specific path becomes \eqref{Probability}:

\begin{equation}\label{Probability2}
p[path] = \frac1Z e^{i\frac{S[path]}{\hbar}}.
\end{equation}

Despite the fact that the probability \eqref{Probability2} is a
complex entity and thus ill-defined in traditional probability
theories it might still have meaning when used to calculate the
entropy. Insert \eqref{Probability2} in \eqref{Entropy2}:

\begin{equation}\label{Entropy3}
\mathcal{H} = -k \sum_{paths} p[path]\Bigg(i\frac{S[path]}{\hbar}
- \log Z \Bigg)
\end{equation}

We also have the normalization \eqref{Normalization}:

\begin{equation}\label{Sum2}
1 = \sum_{paths} p[path],
\end{equation}

and the expression for the expected action \eqref{ExpectedAction}:

\begin{equation}\label{ExpectedAction3}
\langle S \rangle = \sum_{paths} p[path] S[path]
\end{equation}

Together \eqref{Entropy3}, \eqref{Sum2}, \eqref{ExpectedAction3}
and the fact that the partition function $Z$ is path-independent
brings the following general expression for the entropy of the
system:

\begin{equation}\label{EntropyQMGeneral}
\mathcal{H} = - k \Bigg(\sum_{paths} p[path]
i\frac{S[path]}{\hbar} - \sum_{paths} p[path] \log Z \Bigg) = -k
\Bigg(i \frac{\langle S\rangle}{\hbar} - \log Z\Bigg).
\end{equation}

Let us now further assume the special case when the following
identity holds:

\begin{equation}
\psi = Z.
\end{equation}

This identity holds at least when $\psi = \psi(q',t')$ and $t'$ is
a symmetric physical parameter, just as described in section
\ref{PathIntegral}. Let us also assume that the structure of the
wave function is as follows:

\begin{equation}\label{Psi}
\psi = R e^{i\frac{S_c}{\hbar}}
\end{equation}

where $R = |\psi|$ and $S_c$ is the classic action
\cite{Bransden}. This brings the following expression:

\begin{equation}\label{LogPsi}
\log \psi = \log |\psi| + i \frac{S_c}{\hbar}.
\end{equation}

Together \eqref{EntropyQMGeneral} and \eqref{LogPsi} amounts to
the following special case of the entropy:

\begin{equation}
\mathcal{H} = - k \Bigg( i \frac{\langle S\rangle}{\hbar} - i
\frac{S_c}{\hbar} - \log |\psi|\Bigg).
\end{equation}

If we assume the equivalence between the classical action $S_C$
and the expected action $\langle S \rangle$, which is in
accordance with the Ehrenfest theorem \cite{Bransden}, then we get
the following expression for entropy:

\begin{equation}\label{EntropyQM}
\mathcal{H} = k \cdot \log|\psi|.
\end{equation}

An expression similar to \eqref{EntropyQM} was suggested as a
basis for the holographic approach to gravity \cite{Verlinde} in a
somewhat more speculative paper recently \cite{Munkhammar}. In
that approach the constant was suggested to be $k = - 2 k_B$,
where $k_B$ was Boltzmann's constant. The expression for entropy
\eqref{EntropyQM} is strikingly similar to Boltzmann's formula for
entropy in thermodynamics:

\begin{equation}
\mathcal{H} = k_B \cdot \log(W),
\end{equation}

where $\mathcal{H}$ is the entropy of an ideal gas for the number
$W$ of equiprobable microstates \cite{Rief}. The suggested entropy
\eqref{EntropyQM} and it's more general version
\eqref{EntropyQMGeneral} are, up to a scalable constant, measures
of the lack of information in the information-based frame of
reference.

\section{Discussion}
This paper proposes a conceptual foundation for quantum mechanics
based on information brought on by the concept of information
covariance. This approach supports the notion that physics is
largely based on information, a concept that among others Wheeler
strongly endorsed \cite{Wheeler}. The suggested framework in this
paper, building on Lisi's universal action reservoir and Rovelli's
relational quantum mechanics, gives an intuitive description of
physics; Physics in the quantum realm is a consequence of the
incompleteness of information in the local frame of reference. By
setting up an information covariant foundation in the local frame
of reference by means of a maximization of Shannon entropy on the
possible paths of a system we managed to - by using Lisi's theorem
- establish a canonical formulation of relational quantum
mechanics. This implies that Lisi's proposed universal action
reservoir is the inevitable result of the observer ignorance of a
system. We also explicitly calculated the entropy associated with
any quantum mechanical system.


\subsection{Open problems}
This theory primarily serves as a conceptual framework for quantum
mechanics. However it also brings new concepts like for example
the particular entropy \eqref{EntropyQM} of a quantum mechanical
system. The role and the application of the new entropy is not yet
fully investigated. It could, for example, perhaps be related to
holographic theories of gravitation \cite{Munkhammar,Verlinde}. In
addition to this the theory might give interesting effects in
quantum statistical mechanics. Another open issue is how to
construct a general relativistic approach to this theory. Such a
theory ought to arrive at some similar problems that many quantum
gravity theories have stumbled upon because this theory is merely
a relational version of the canonical path integral formulation of
quantum mechanics.

\subsection{Final comments}
When cast in a local frame of reference physics has only a limited
amount of information with which to function. When physics is
fundamentally bound by a limited amount of information
probabilistic effects will occur. By maximizing the entropy the
probabilistic effects quantum physics arises. The result of
subjecting classical mechanics to incomplete information is
quantum mechanics.

%
%
%
%
%


%




%
%
%
%

\section{Acknowledgments}
This work would not have been possible without the inspiration and
the valuable ideas shared in the great works of G. Lisi, C.
Rovelli and J-W. Lee. I would also like to thank prof. Lee for a
review of the paper and giving valuable comments.


\end{document}